\documentclass[prl,twocolumn,showpacs,amssymb]{revtex4}
\newcommand{\be}{\begin{equation}}
\newcommand{\ee}{\end{equation}}
\newcommand{\bea}{\begin{eqnarray}}
\newcommand{\eea}{\end{eqnarray}}
\usepackage{graphicx}
\usepackage{amsmath}

\begin{document}

\title{BCS theory of time-reversal-symmetric Hofstadter-Hubbard model}
\author{R. O. Umucal\i lar$^{1,2}$ and M. Iskin$^{1}$}
\affiliation{$^{1}$Department of Physics, Ko\c{c} University, Rumelifeneri Yolu, 34450 Sar\i yer, 
Istanbul, Turkey\\$^{2}$Department of Physics, Mimar Sinan Fine Arts University, 34380 \c{S}i\c{s}li, 
Istanbul, Turkey}

\begin{abstract}
The competition between the length scales associated with the periodicity of a 
lattice potential and the cyclotron radius of a uniform magnetic field is known to
have dramatic effects on the single-particle properties of a quantum particle, 
e.g., the fractal spectrum is known as the Hofstadter butterfly. Having this intricate
competition in mind, we consider a two-component Fermi gas on a square optical 
lattice with opposite synthetic magnetic fields for the components, and study its 
effects on the many-body BCS-pairing phenomenon. By a careful addressing of 
the distinct superfluid transitions from the semi-metal, quantum spin-Hall insulator 
or normal phases, we explore the low-temperature phase diagrams of the model, 
displaying lobe structures that are reminiscent of the well-known Mott-insulator 
transitions of the Bose-Hubbard model.
\end{abstract}
\maketitle

{\it Introduction.---}
The discoveries of integer and fractional quantum-Hall effects in 1980s brought 
a new breath in solid-state and condensed-matter physics, attracting 
a never-ending interdisciplinary attention since then~\cite{QHE}. 
For instance, some mathematical ideas from the topology turned out to be very 
successful in explaining the robustness of these effects, making the so-called 
topological insulators a very popular research theme in modern physics. 
These materials are intrinsically insulating in the bulk but have 
conducting edge/surface states that are robust against local perturbations~\cite{TI review}. 
While the earlier proposals require a broken $\mathcal{T}$ symmetry as can be 
realized under an external magnetic field, the quantum spin-Hall insulators (SHI) 
preserved it~\cite{Kane, QSH experiment} in such a way that the currents carried 
by electrons with different spin states flow in opposite directions along the edges 
of the sample without dissipation.

In the mean time, the successful creation of atomic BECs in 1990s and the 
tunable BCS-BEC crossover in 2000s ignited researchers to transfer many of the 
model Hamiltonians developed in physics all across-the-board into the realm 
of ultra-cold atomic systems~\cite{Cold atoms}. For instance, the recent production
of synthetic magnetic fields for neutral atoms~\cite{Artificial} was followed by 
the realization of the celebrated Harper-Hofstadter model~\cite{Harper,Hofstadter} 
in the presence of an optical lattice~\cite{Hofstadter experiment 1,
Hofstadter experiment 2, ketterle13, ketterle15, goldman15, spielman15, greiner16}, 
the $\mathcal{T}$-preserving schemes of which were also developed to realize 
the quantum spin-Hall Hamiltonians~\cite{Hofstadter experiment 1, ketterle13}.
Besides having an intriguing fractal spectrum, the generic Harper-Hofstadter model 
not only features the integer quantum-Hall effect~\cite{Thouless} but it also hosts 
Dirac-cone physics for certain magnetic fluxes. These cones are at the heart of 
certain phenomena in graphene-type materials with a honeycomb lattice, including 
the quantum spin-Hall effect with an additional gap-opening mechanism~\cite{Kane 2} 
and the semi-metal (SM)-BCS superconductivity transition~\cite{Paramekanti, Sonin}.

\begin{figure}[htbp]
\includegraphics[scale=0.35]{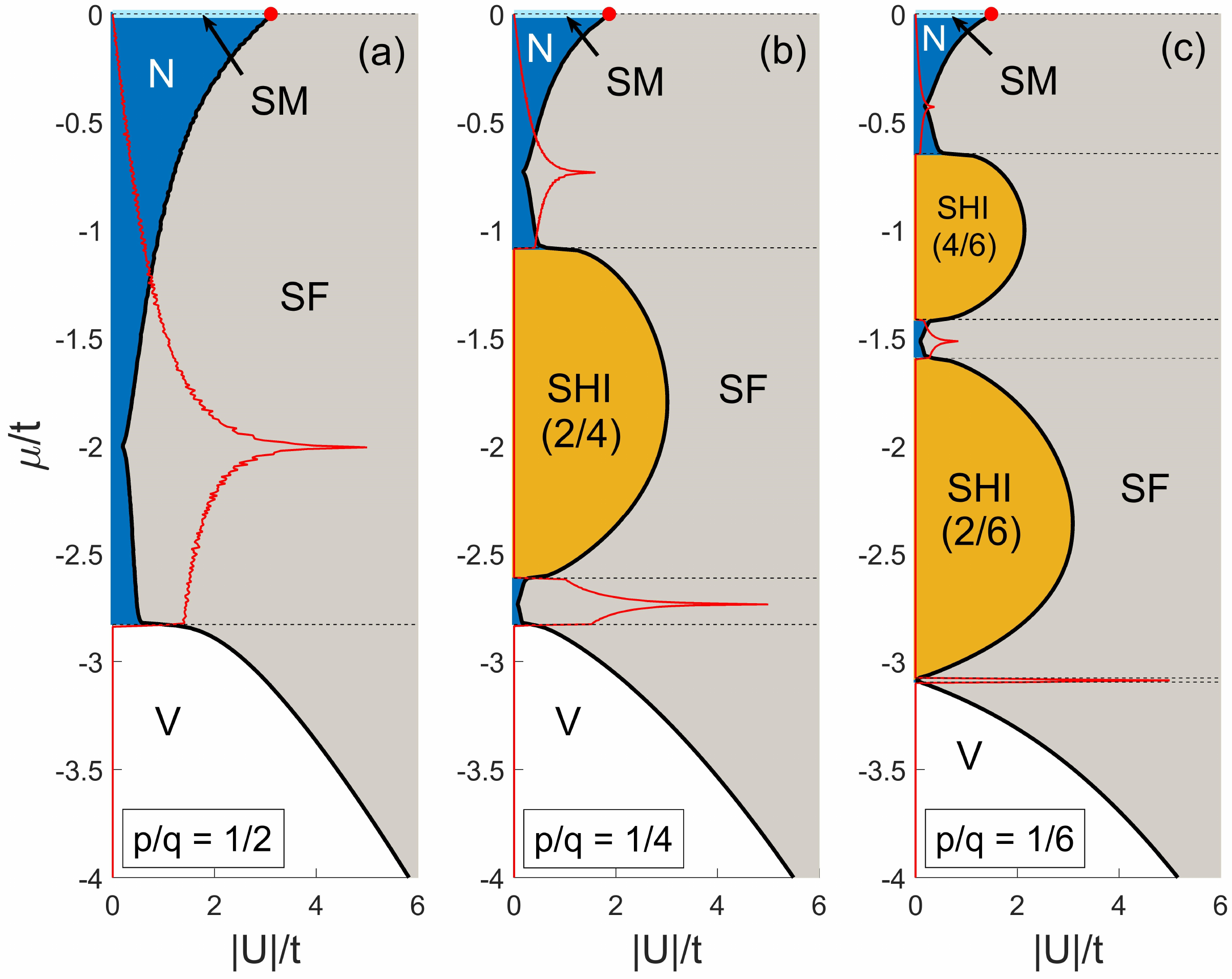}
\caption{(Color online) 
Phase diagrams for $\alpha = p/q$ with $p = 1$ and even-$q$ denominators. 
Critical interaction strength $|U_c|/t$ is shown as a function of $\mu/t$ by a thick 
black curve separating V (particle vacuum), N (normal), SM (semi-metal), 
quantum SHI (spin-Hall insulator), and SF (superfluid) phases at $k_BT = 10^{-4}t$. 
Total filling $F$ of an SHI phase is shown in parentheses. Red dot at $\mu =0$ shows 
the triple point at which N, SM and SF phases meet. Density of states $D(\varepsilon)$ 
is displayed in arbitrary units by a thin red curve ($\varepsilon$ and $\mu$ axes coincide). 
Horizontal dashed lines mark the band edges including $\varepsilon = 0$.
\label{p1q246Uc}
}
\end{figure}

The Hofstadter physics is enriched even more by the effects of interactions. 
For instance, while it gives rise to a complicated shape for the transition 
boundary between the superfluid (SF) and Mott-insulator phases in the context of 
repulsive Bose-Hubbard model~\cite{BoseHubbard, BosonicHH}, it promotes 
a playground for a variety of inhomogeneous SF phases with vortex-lattice or 
stripe orders in the context of attractive Hubbard model~\cite{Zhai, Iskin, umucalilar16}. 
Furthermore, the $\mathcal{T}$-preserving Hofstadter-Hubbard models have 
also been investigated soon after its reazilations with cold 
atoms~\cite{Hofstadter experiment 1, ketterle13}, offering a wealth of phase transitions. 
For instance, while the repulsive interactions may pair up the fractional quantum-Hall 
states to form fractional quantum spin-Hall states~\cite{Ueda} in a Bose gas, 
it may drive a phase transition from an SM to an antiferromagnetic 
insulator in a Fermi gas~\cite{Hofstetter}. In addition, the attractive interactions
in a Fermi gas may drive phase transitions from an SM or an SHI to an SF 
phase~\cite{Troyer, Torma, Nishida}, motivating this work.

In this paper, we present a systematic analysis of the SF transition in a 
$\mathcal{T}$-symmetric Hofstadter-Hubbard model, and explore the 
resultant phase diagrams for a wide-range of model parameters. 
Unlike the existing literature on the Hofstadter-Hubbard model with a 
broken $\mathcal{T}$-symmetry~\cite{Zhai, Iskin, umucalilar16}, 
we first show that the SF phase is characterized by a spatially-uniform 
order parameter, despite the presence of a complicated single-particle 
spectrum brought about by the interplay of the lattice potential and the 
magnetic field. Based on this observation, we identify distinct transitions 
from the SM, SHI, normal (N), and vacuum (V) phases to the homogenous 
SF phase with respect to the single-particle density of states of the 
multi-band energy spectrum. See Fig.~\ref{p1q246Uc} for a typical illustration,
whose lobe structures are reminiscent of the Mott-insulator transitions 
of the SF Bose gas on a lattice.
We also pay special attention to the magnetic-flux dependence of the interaction 
threshold for the SM-SF triple point and determine the hallmark attributes 
of the SF order parameter depending on the type of the transition, 
providing analytical expressions in various limits. Besides capturing the 
essential physics of the model Hamiltonian, we hope that our simpler 
mean-field results may also work as a benchmark for more accurate 
but numerically-demanding QMC simulations~\cite{Troyer}.

{\it Mean-field theory.---} 
To describe the kinematics of a quantum particle in a tight-binding square lattice 
potential, we start with the following single-particle Hamiltonian, 
$
H_B = - t\sum_{\langle ij\rangle \sigma}(e^{\mathrm{i}2\pi \phi_{ij}^{\sigma}} 
c^\dag_{i\sigma}c_{j\sigma}+\text{h.c.}), 
\label{eq:Hofstadter Hamiltonian}
$
where $t > 0$ is the hopping amplitude between nearest-neighbor sites 
$\langle ij\rangle$, i.e., $c^{\dagger}_{i\sigma}$ ($c_{i\sigma}$) creates (annihilates) 
a fermion at site $i \equiv (i_x, i_y)$ with pseudo-spin index 
$\sigma\equiv\{\uparrow,\downarrow\}$, and \text{h.c.} is the Hermitian conjugate. 
The perpendicular magnetic field is taken into account via minimal coupling, 
where the hopping particle acquires a spin-dependent phase,
$
\phi_{ij}^{\sigma} = (s_\sigma/\phi_0) \int_{{\bf r}_j}^{{\bf r}_i} {\bf A}\cdot d{\bf r},
$ 
with $s_\uparrow = +1$ and $s_\downarrow = -1$. Here, $\phi_0$ is the magnetic-flux 
quantum and ${\bf A} = (0, Bx)$ is the vector potential in the Landau gauge with 
$B$ the magnitude of the effective magnetic field. For a given magnetic-flux quanta 
per unit cell $\alpha = Ba^2/\phi_0$, with $a \to 1$ the lattice constant, 
the particle gains an Aharonov-Bohm factor $\exp(\mathrm{i} 2\pi s_\sigma \alpha)$ 
after traversing a loop around the unit cell. This is the so-called 
{\it time-reversal-symmetric} Hofstadter model~\cite{Troyer, Hofstetter} as 
realized in recent cold-atom experiments~\cite{Hofstadter experiment 1, ketterle13}.
When $\alpha$ is a rational fraction $p/q$ with $p$ and $q$ relatively-prime integers,
the spectrum for each spin state consists of $q$ subbands which split from the 
tight-binding $s$-band of the field-free case. Energy versus $\alpha$ diagram has
a fractal structure and is usually called the Hofstadter butterfly~\cite{Hofstadter}.

Under the validity of this model, the non-interacting Fermi gas is described by 
$
H_0 = H_B - \mu \sum_{i\sigma} n_{i\sigma},
$
where 
$
n_{i\sigma} = c^{\dagger}_{i\sigma}c_{i\sigma}
$ 
is the number operator and $\mu$ is the common chemical potential for both spin 
states. Furthermore, having short-ranged attractive interactions in mind, we 
adopt a BCS-like mean-field approximation for pairing, and consider an 
on-site term,
$
H_I = - \sum_{i}(\Delta_i c^\dag_{i\uparrow}c^\dag_{i\downarrow}+\text{h.c.})
- \sum_{i}|\Delta_i|^2/U,
$
where
$
\Delta_i = U \langle c_{i\uparrow} c_{i\downarrow} \rangle
$
is the SF order parameter. Here, $U \le 0$ and $\langle \ldots \rangle$ denotes 
the thermal average. Next, we switch to the momentum-space representation, 
and define $n = 1,\ldots,q$ band operators $d_{{\bf k}n\sigma}$ in terms of the 
Fourier-expansion coefficients of $c_{i\sigma}$, i.e., 
$
c_{{\bf k}\beta\sigma} = \sum_n g^{n}_{\beta\sigma}({\bf k})d_{{\bf k}n\sigma},
$
where $g^{n}_{\beta\sigma}({\bf k})$ is the $\beta$th component of the $n$th 
eigenvector of the single-particle problem with energy $\varepsilon_{{\bf k}n\sigma}$.
Since the magnetic field imposes a new translational symmetry and enlarges 
the unit cell by a factor of $q$ in the $x$ direction, the Brillouin zone is reduced to 
${\bf k}_x \in [-\pi/q,\pi/q)$ and ${\bf k}_y \in [-\pi,\pi)$~\cite{Hofstadter,Kohmoto},
and we label inequivalent sites in the enlarged cell by $\beta = 0,\ldots,q-1$, 
where $i_x = sq+\beta$ with $s$ locating the supercell. 
The total $\mathbf{k}$-space Hamiltonian $H_{MF} = H_0+H_I$ can be written as
\bea
H_{MF} &=& \sum_{n{\bf k}\sigma}\epsilon_{{\bf
k}n\sigma}d^\dag_{{\bf k}n\sigma} d_{{\bf k}n\sigma}
- \frac{M}{qU} \sum\limits_{l\beta} |\Delta^l_\beta|^2 \\
&-& \sum\limits_{l\beta n n^\prime{\bf k}}\left[\Delta^l_{\beta}
g^{n *}_{\beta\uparrow}({\bf k}^{l}_{+})g^{n^\prime *}_{\beta\downarrow}({\bf k}^{l}_{-}) 
d^\dag_{{\bf k}^{l}_{+}n\uparrow}d^\dag_{{\bf k}^{l}_{-}n^\prime\downarrow}
+\text{h.c.}\right] \nonumber,
\label{eq:MF Hamiltonian}
\eea
where $\epsilon_{{\bf k}n\sigma}=\varepsilon_{{\bf k}n\sigma}-\mu$, and
the $q \times q$ order parameter set
$
\Delta^{l}_\beta = -(qU/M)\sum_{n n^\prime{\bf k}}
g^n_{\beta\uparrow}({\bf k}^{l}_{+})g^{n^\prime}_{\beta\downarrow}({\bf k}^{l}_{-})\langle 
d_{{\bf k}^{l}_{-}n^\prime\downarrow}d_{{\bf k}^{l}_{+}n\uparrow}\rangle
$
determines $\Delta_i$ through 
$
\Delta_i = \sum_{l} \Delta^{l}_\beta e^{\mathrm{i} (Q_{lx} s + Q_{ly} i_y)}.
$
Here, $M$ is the total number of lattice sites and ${\bf k}^{l}_{\pm} = \pm{\bf k}+{\bf Q}_l/2$ 
with ${\bf Q}_l = (Q_{lx}, Q_{ly})$ the center-of-mass momentum of Cooper pairs.
As $\varepsilon_{{\bf k}n\sigma}$ is $q$-fold degenerate in any given band for momenta 
${\bf k}$ and ${\bf k}+{\bf K}_l$ with ${\bf K}_l\equiv \{(0, 2\pi l p/q)\}$ and $l =0,\dots,q-1$, 
we treat pairing with ${\bf Q}_l \equiv {\bf K}_l$ on equal footing~\cite{Zhai}.

By solving these self-consistency equations for a wide range of parameters, 
we confirm that the thermodynamic potential is minimized by the solution 
$\Delta^{l}_\beta = \Delta \delta_{l0}$ with $\delta_{ij}$ the Kronecker-delta, 
in such a way that a single order parameter 
$
\Delta = -(U/M)\sum_{n{\bf k}}\langle d_{{\bf k}n\downarrow}d_{-{\bf k}n\uparrow}\rangle
$ 
characterizes the resultant homogenous SF phase~\cite{BdG}. 
This is unlike the usual Hofstadter-Hubbard model with a broken $\mathcal{T}$ 
symmetry, where inhomogenous SF phases require a non-trivial set of $q \times q$ 
parameters, e.g., a vortex-lattice solution~\cite{Zhai, umucalilar16}. The large $|U|/t$ 
limit is particularly illuminating beyond which the entire Fermi gas consists of 
many-body bound states that eventually form two-body bound states, i.e., bosonic 
molecules, experiencing no net magnetic field. Thus, thanks to the $\mathcal{T}$ 
symmetry of the present model, the order parameter equation simplifies to
\bea
\dfrac{1}{U} = -\dfrac{1}{M}\sum_{n{\bf k}}\dfrac{1}{2E_{{\bf k}n}} 
\tanh\left(\frac{E_{{\bf k}n}}{2k_BT}\right),
\label{eq:gap equation}
\eea
where 
$
E_{{\bf k}n} = \sqrt{\epsilon_{{\bf k}n}^2+\Delta^2}
$ 
is the quasiparticle energy in a given band $n$ as the dispersion $\varepsilon_{{\bf k}n}$ 
is equal for both spin states, $\Delta$ is taken as a real parameter without loosing 
generality, $k_B$ is the Boltzmann constant, and $T$ is the temperature. 
Here, $\mu$ is determined by the total number of particles 
$
N = \sum_{i\sigma} \langle n_{i\sigma}\rangle,
$ 
leading to the number equation
\bea
F = \dfrac{1}{qM}\sum_{n{\bf k}}\left[1-\dfrac{\epsilon_{{\bf k} n}}{E_{{\bf k}n}}
\tanh\left(\frac{E_{{\bf k}n}}{2k_BT}\right)\right],
\label{eq:number equation}
\eea
where $0 \le F = N/M \le 2$ is the total particle filling. 
While we recover the familiar expressions
$
\Delta = (|U|/2-4t^2/|U|) \sqrt{F(2-F)}
$
and
$
\mu = -(|U|/2-8t^2/|U|)(1-F)
$
for the bosonic molecules in the strong-coupling limit when $\Delta\gg t$ or
equivalently $|U| \gg t$, the weak-coupling limit turns out to be quite rich showing 
a variety of distinct phases and transitions in between. Next, we construct typical 
phase diagrams for $\mu \le 0$, as the solutions are mirror-symmetric 
around $\mu = 0$ or equivalently $F = 1$ due to the particle-hole symmetry 
of the parent Hamiltonian.

\begin{figure}[htbp]
\includegraphics[scale=0.35]{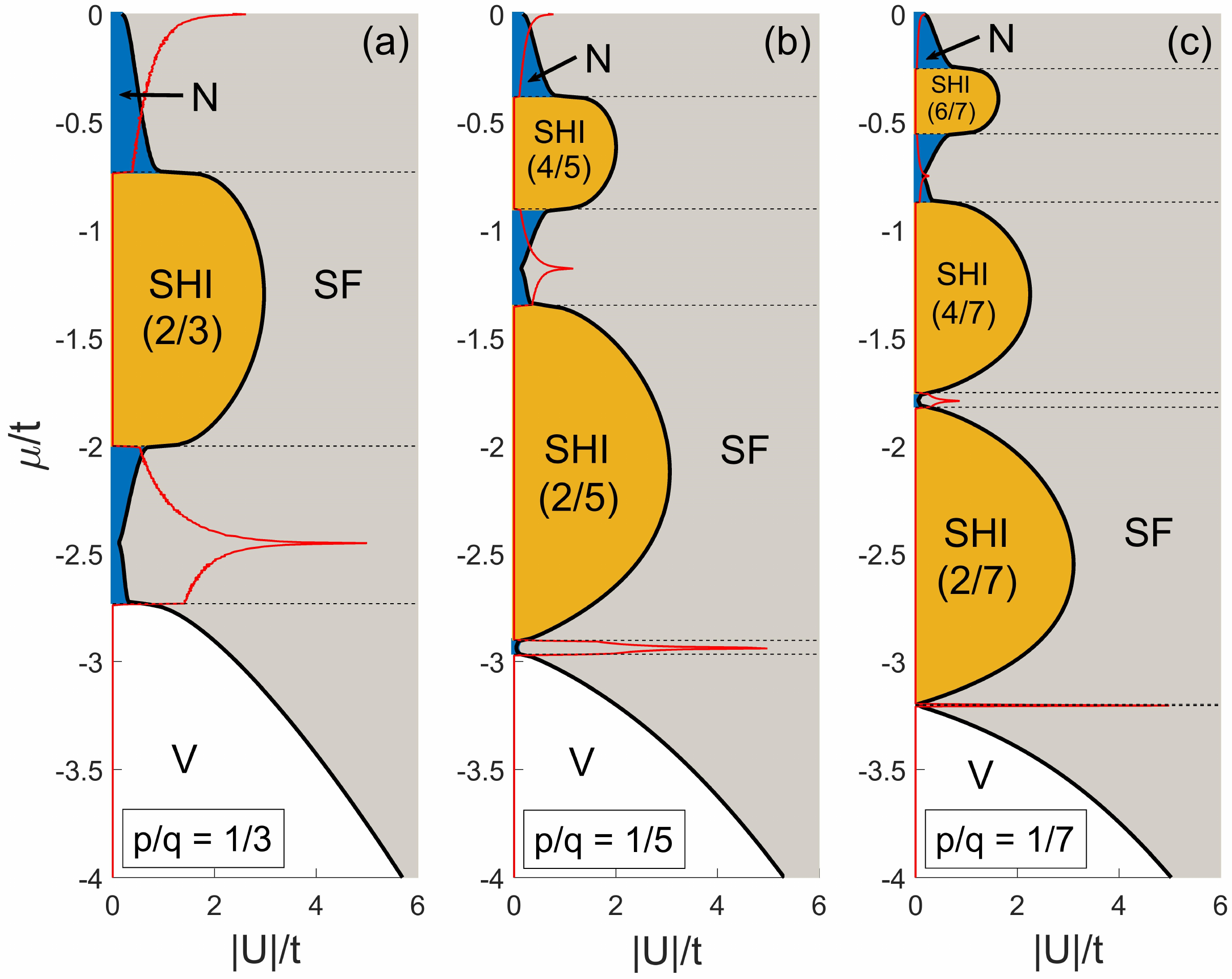}
\caption{(Color online)
Similar to Fig.~\ref{p1q246Uc} but for odd-$q$ denominators.
\label{p1q357Uc}
}
\end{figure}

{\it Low-temperature phase diagrams.---} 
The critical interaction threshold $|U_c|$ above which the system develops SF 
correlations with $\Delta \ne 0$ can be obtained by setting $\Delta \to 0$ in 
Eqs.~(\ref{eq:gap equation}) and (\ref{eq:number equation}). For instance, 
our low-$T$ phase diagrams that are shown in Figs.~\ref{p1q246Uc} and 
\ref{p1q357Uc} reveal four distinct SF transitions:
({\it i}) an SM-SF transition when $\mu=0$ or $F = 1$ for even $q$, 
({\it ii}) an SHI-SF transition when $\mu$ lies within a band gap or $F = 2s/q$ 
with $s \le q$ an integer,
({\it iii}) an N-SF transition when $\mu$ lies within a band or $F \ne 2s/q$, and
({\it iv}) a V-SF transition when $\mu$ lies below the lowest available band 
or $F \to 0$. 
We note that even though the mean-field framework is known to be less-accurate 
in two dimensions, not to mention the critical role played by the multi-band 
spectrum, it should be considered as a qualitative description of the system 
at the best-case scenario. Furthermore, given its semi-analytic nature, it not 
only helps us build the intuition behind these competing phases but it also 
serves as an ultimate benchmark for fully-numerical QMC simulations~\cite{Troyer}. 

\begin{figure}[htbp]
\includegraphics[scale=0.49]{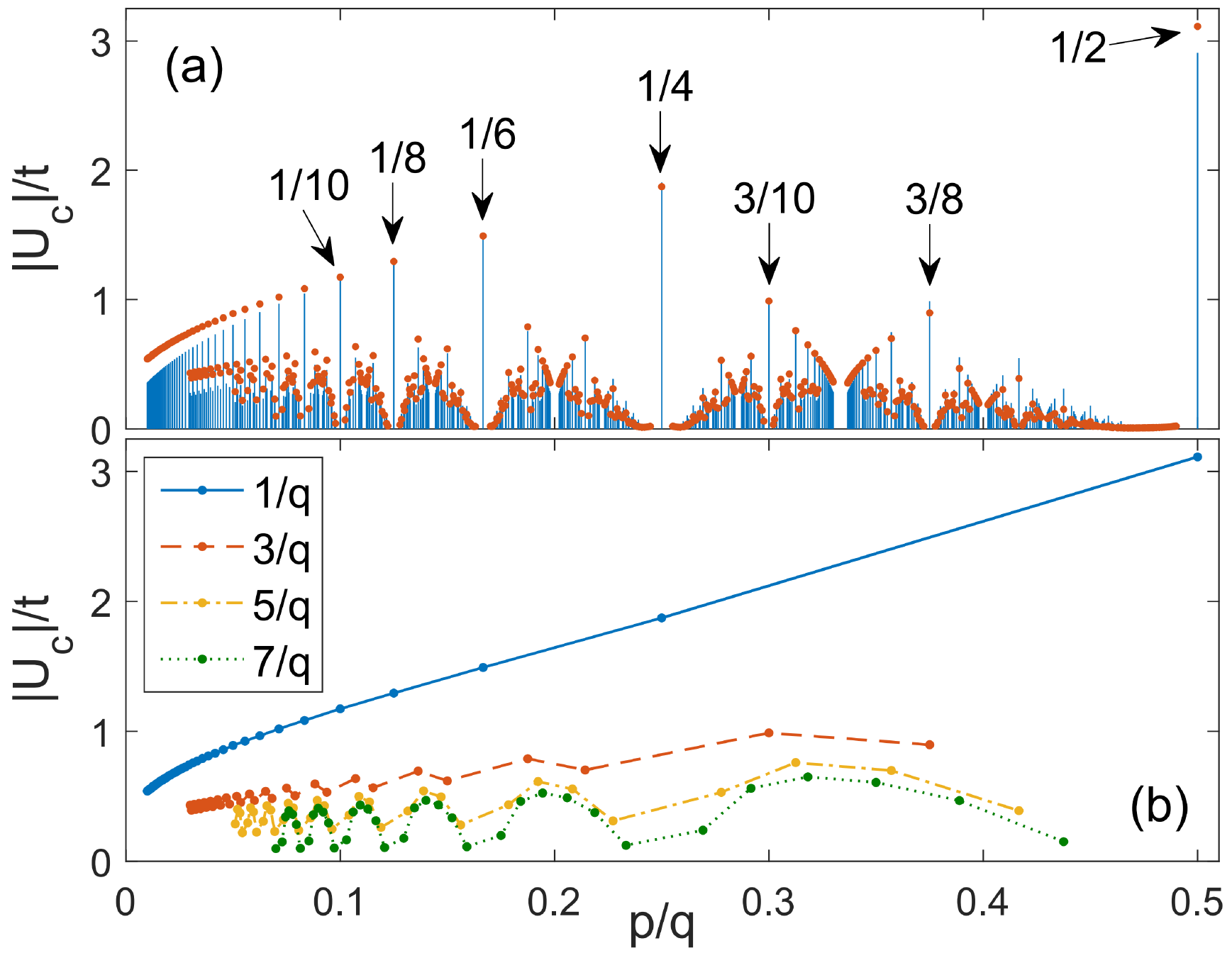}
\caption{(Color online)
SM-SF transition thresholds at $k_BT = 10^{-4}t$ for all unique 
$\alpha = p/q \leq 1/2$ ratios with even-$q$ denominators up to 100. 
(a) While the red points are our self-consistent solutions, the vertical blue lines 
correspond to $(2W/t)^{0.44}$ where $W$ is the total bandwidth of the 2-central 
bands for a given $\alpha$. In (b) the same data is grouped according to the 
numerator $p$, where the connecting lines are drawn as a guide, 
showing the oscillatory dependence.
\label{UcvsAlpha}
}
\end{figure}

First of all, the SM-SF transition at $F = 1$ (i.e., half-filling) for even-$q$ 
denominators is caused by the presence of $q$ linearly-dispersing Dirac 
cones in the magnetic Brillouin Zone. It turns out that even though 
$\Delta \to 0$ at the transition boundary, the Dirac cones guarantee a small 
energy window of $\mathbf{k}$-space region around $\varepsilon = 0$ with
$\Delta > |\varepsilon|$, no matter how small the energy window is. This leads
to a finite triple point $U_c \ne 0$ as shown in Fig.~\ref{p1q246Uc}. 
The $T = 0$ limit of $U_c$ is determined by 
$
M/U_{c} = -\sum_{n{\bf k}}1/(2|\varepsilon_{{\bf k}n}|),
$
near which
$
\Delta_0 = (U_{c}-U)/(CU_{c}^2)
$
increases linearly with $|U|$, where $D(\varepsilon) = C|\varepsilon|$ is the 
low-energy density of states near the cones. 
At $T = 0$, we find $|U_c|/t \approx 3.111$ and $1.871$ for $q = 2$ and $4$, 
respectively, and the complicated dependence of $U_c$ on $q$ is closely 
related to the band width of the central bands as shown in Fig.~\ref{UcvsAlpha}(a) 
for $q$ up to 100. For a given prime number $p$, the oscillatory dependence 
shown in Fig.~\ref{UcvsAlpha}(b) is a consequence of the self-similar fractal 
spectrum in such a way that each $\alpha$ interval between $1/q$ and 
$1/(q+2)$ contains $p-1$ data points with distinct $p/q$ ratios.

Away from half filling, when the filling fraction is $F=2s/q$ with $s\leq q$ an 
integer, there are $s$ fully-occupied bands and the system is an SHI. 
At $T = 0$, the SHI-SF transition boundary is determined by 
$
M/U_{c} = -\sum_{n{\bf k}}1/(2|\epsilon_{{\bf k}n}|),
$
near which
$
\Delta_0 = \sqrt{(U_{c}-U)/(C_0 U_{c}^2)}
$
increases as a square-root with $|U|$, where
$
M C_0 = \sum_{n{\bf k}} 1/(4|\epsilon_{{\bf k}n}|^3)
$
is a constant for a given $q$. On the other hand, since $F \ne 2s/q$ corresponds 
to an N phase with a partially-occupied band, we find that $U_c \to 0$ as 
$T \to 0$~\cite{supmat}, near which
$
\Delta_0 = 2|\mu|\exp\{1/[D(\mu)U]-1\}
$
increases exponentially with $|U|$ for even-$q$ values close to the half filling 
when $\Delta_0 \ll |\mu| \approx 0$. The BCS-like $D(\varepsilon)$ dependence is 
clearly seen in Figs.~\ref{p1q246Uc} and \ref{p1q357Uc}, where, while $T \ne 0$ 
causes $U_c \ne 0$ in general, its magnitude is inversely related to $D(\varepsilon)$.
Note that, since the total band width is constrained by $8t$ in the $q \to \infty$ limit, 
increasing $q$ flattens the band widths of each of the $q$ bands, leading to 
a singular $D(\varepsilon)$ with discrete structure. As the N regions shrink and 
become hardly visible even at $T \ne 0$, our large-$q$ phase diagrams~\cite{supmat} 
are reminiscent of the Mott-insulator transitions of the SF Bose gas on 
a lattice~\cite{Cold atoms, Fisher89}. 

\begin{figure}[htbp]
\includegraphics[scale=0.5]{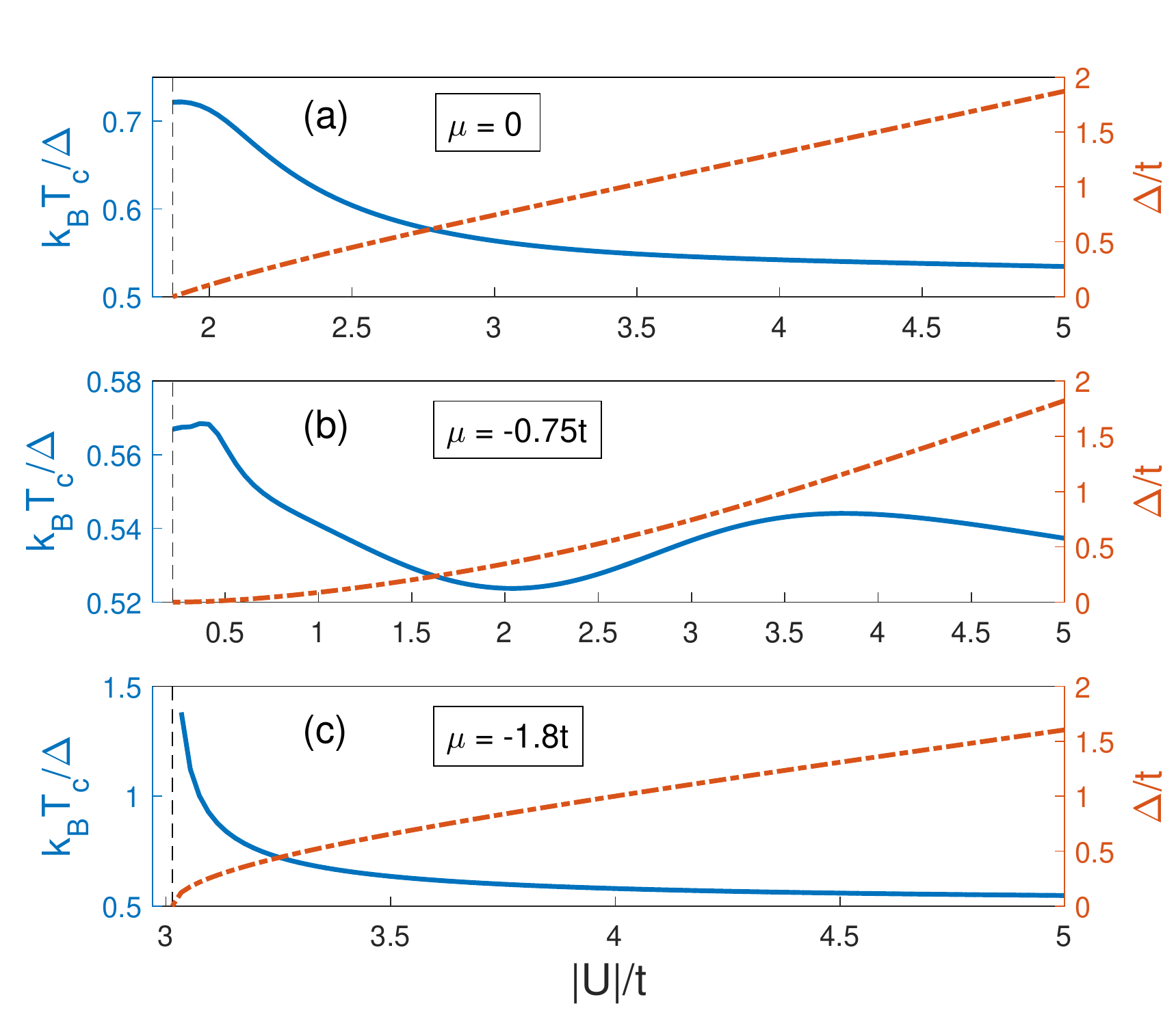}
\caption{Color online)
(Left) Ratios of the critical temperature $k_B T_c$ and $\Delta$ at $k_BT = 10^{-4}t$ 
for $\alpha = 1/4$. (Right) $\Delta/t$ as dash-dotted red lines. 
Vertical dashed lines correspond to the $|U_c|/t$ thresholds for the 
(a) SM-SF,
(b) N-SF, and 
(c) SHI-SF transitions discussed in the text.
\label{TcDelta_vs_U_p1q4}
}
\end{figure}

This low-$T$ analysis clearly show that, depending on whether $\Delta_0$ increases 
as a linear, square-root or exponential function of $|U|$, one can distinctly characterize 
the corresponding type of the SF transition. For instance, we illustrate $\Delta$ for all
3 types on the right axis of Fig.~\ref{TcDelta_vs_U_p1q4}, where we set $\alpha = 1/4$ 
and $k_BT = 10^{-4}t$. To further support this finding, we also show the ratio 
$k_BT_c/\Delta$ on the left axis of the same figure, where $T_c$ is the corresponding
SF transition temperature. Setting $\Delta = 0$ and $\mu = 0$ in Eq.~(\ref{eq:gap equation}), 
and assuming $T_c$ is small, we find
$
k_BT_c = (U_{c}-U)/(2\ln2 CU_{c}^2)
$
as $U \to U_c$ in the weak-coupling limit near the $T = 0$ SM-SF transition boundary.
This suggests 
$
k_BT_c/\Delta_0 = 1/(2\ln2) \approx 0.721
$
in the $U \to U_c$ limit, which is in perfect agreement with 
Fig.~\ref{TcDelta_vs_U_p1q4}(a).
Similarly, setting $\Delta = 0$ and $\mu \approx 0$ in Eq.~(\ref{eq:gap equation}), 
and assuming $k_B T_c \ll |\mu|$, we find
$
k_BT_c = (2|\mu|/\pi)\exp\{1/[D(\mu)U]+\gamma-1\}
$
for even-$q$ values as $U \to 0$ in the weak-coupling limit near the N-SF 
transition boundary, where $\gamma \approx 0.577$ is the Euler's constant.
This suggests 
$
k_BT_c/\Delta_0 = e^\gamma/\pi \approx 0.567
$
in the $U \to U_c = 0$ limit, which is again in perfect agreement with 
Fig.~\ref{TcDelta_vs_U_p1q4}(b).
Lastly, setting $\Delta = 0$ in Eq.~(\ref{eq:gap equation}), and assuming $\mu$ is in the 
middle of one of the band gaps and $k_B T_c \ll A$ is small, we find
$
k_B T_c \sim -A/\{2\ln[Aq(U_c-U)/(4U_c^2)]\}
$
as $U \to U_c$ in the weak-coupling limit near the $T = 0$ SHI-SF transition boundary, 
where $A$ is approximately the corresponding band gap between the highest-occupied 
and lowest-unoccupied band. This suggests 
$
k_B T_c/\Delta_0 \to \infty
$
in the $U \to U_c$ limit, i.e., a logarithmic divergence of the form 
$\lim_{x \to 0} 1/|\sqrt{x}\ln x|$, which is also in perfect agreement with 
Fig.~\ref{TcDelta_vs_U_p1q4}(c). 
In the strong-coupling limit when $|U| \gg t$, it is well-known that the mean-field 
$T_c$ is solely related to the formation of pairs and it has nothing to do with 
the actual SF transition. Setting $\Delta = 0$ in Eq.~(\ref{eq:gap equation}), 
and assuming $k_BT_c \gg |\mu|$ leads to
$
k_B T_c = |U|/4
$
which is the case around $\mu \approx 0$ (or half filling $F \approx 1$), while assuming 
$k_BT_c \ll |\mu|$ leads to
$
k_B T_c = -|\mu|/\ln (F/2)
$
and $|\mu| \approx |U|/2$, which is the case for low filling $F \ll 1$. 
Therefore, in the $|U| \gg t$ limit, while all 3 of our numerical results approach
$
k_B T_c/\Delta_0 \to 0.5
$
in Fig.~\ref{TcDelta_vs_U_p1q4}, we expect a diverging
$
k_B T_c/\Delta_0 \simeq -1/[\sqrt{2F} \ln (F/2)]
$ 
ratio as $F \to 0$. It is notable that this logarithmic divergence is quite similar 
in structure to that of the weak-coupling one near the SHI-SF transition 
discussed just above.

Furthermore, in the weak-coupling limit near the SF transition boundary, using 
$\Delta \ll k_B T_c$ near $T_c$, we find 
$
\Delta = \sqrt{8\ln 2} k_B T_c \sqrt{1-T/T_c} \approx 1.698 \Delta_0 \sqrt{1-T/T_c}
$
for even-$q$ values when $\mu = 0$ is at half filling,
$
\Delta = \sqrt{8 \pi^2/[7\xi(3)]} k_B T_c \sqrt{1-T/T_c} \approx 1.736 \Delta_0 \sqrt{1-T/T_c}
$
for even-$q$ values when $\mu \approx 0$ is in the band, and
$
\Delta = \{[1/(M C_0 k_B T_c)] \sum_{n {\bf k}}e^{-|\epsilon_{{\bf k}n}|/(k_B T_c)}\}^{1/2}
\sqrt{1-T/T_c}
\sim \sqrt{A/(2k_B T_c)} \Delta_0 \sqrt{1-T/T_c}
$
when $\mu$ is in the band gap with $A \gg k_B T_c$. 
Lastly, in the large-$q$ limit when $|\epsilon_{\mathbf{k} n}| \ll k_B T_c$ 
for any one of the bands, we find
$
\Delta = \sqrt{12} k_B T_c \sqrt{1-T/T_c} = \sqrt{3} \Delta_0 \sqrt{1-T/T_c},
$
the coefficient $\sqrt{3} \approx 1.732$ of which almost coincides with 
that of the BCS expression. Therefore, the $T$-dependences of $\Delta$ 
are all alike near $T_c$ up to the prefactor, in the characteristic form of 
a second-order phase transition. 

We end this paper by noting that, in addition to the recent 
proposals for distinguishing different SHI lobes~\cite{ketterle13, Troyer}, 
the SM-SF and SHI-SF transitions may be directly probed by measuring 
the density profiles, and studying the resultant `wedding cake' 
structures~\cite{Cold atoms}. In addition, the SHI phases can be further 
identified by the density profiles via an effective `Hall conductance'  introduced 
through the well-known Streda formula~\cite{Streda-density}.

{\it Conclusions.---}
In summary, we used the $\mathcal{T}$-symmetric Hofstadter-Hubbard model on a 
square optical lattice in order to describe and study the BCS-pairing correlations
of a two-component Fermi gas that is experiencing opposite synthetic magnetic 
fields for its components. We found rich phase diagrams involving distinct 
SF transitions from the SM, quantum SHI or N phases, the lobe structures 
of which are reminiscent of the Mott-insulator transitions of the SF Bose gas 
on a lattice~\cite{Cold atoms, Fisher89}. Given the ongoing cold-atom experiments 
in simulating such models~\cite{Hofstadter experiment 1, ketterle13}, there is 
no doubt that even though our mean-field description may only capture qualitative 
physics of the model Hamiltonian in two dimensions, which is further complicated
by the multi-band spectrum, it not only offers a less-accurate but analytically-tractable 
analysis in helping us shape the intuition behind the competing phases but it 
also paves the way as an ultimate benchmark for more-accurate yet 
fully-numerical QMC simulations~\cite{Troyer}.

\begin{acknowledgments}
M. I. acknowledges funding from T{\"U}B{\.I}TAK Grant No. 1001-114F232 
and the BAGEP award of the Turkish Science Academy. 
\end{acknowledgments}

\newpage
\begin{widetext}
\section{Supplementary Online Material}

\begin{figure}[htbp]
\includegraphics[scale=0.4]{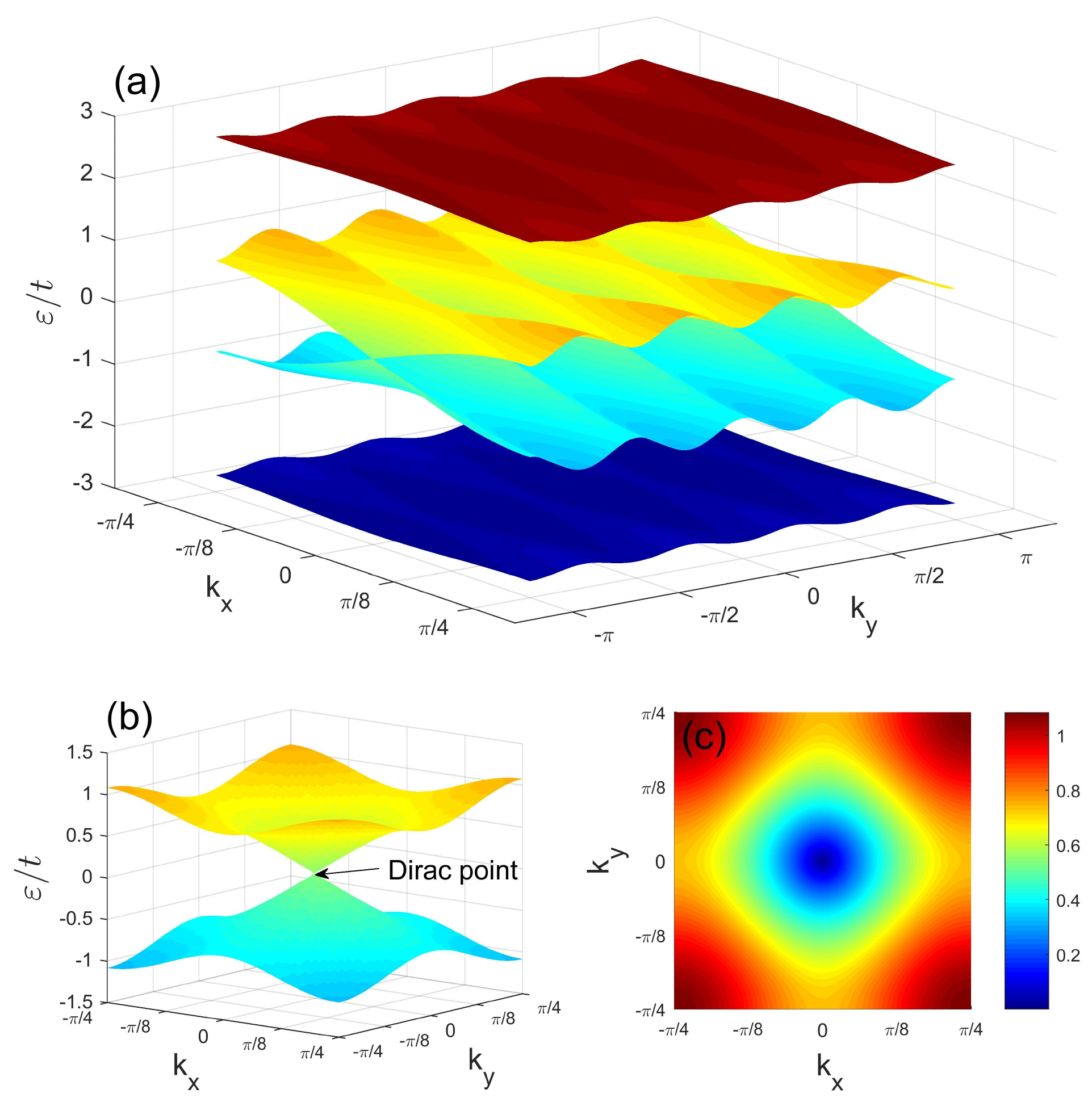}
\caption{Color online)
(a) Energy spectrum of the Hofstadter model for $\alpha = p/q$ with $p=1$ and $q=4$ 
in the first magnetic Brillouin zone. The $\mathbf{k}$-space structure of one of the $q$ Dirac 
cones is shown with greater details in (b) and (c).
\label{p1q1mu0gcvsh}
}
\end{figure}
\begin{figure}[htbp]
\includegraphics[scale=0.06]{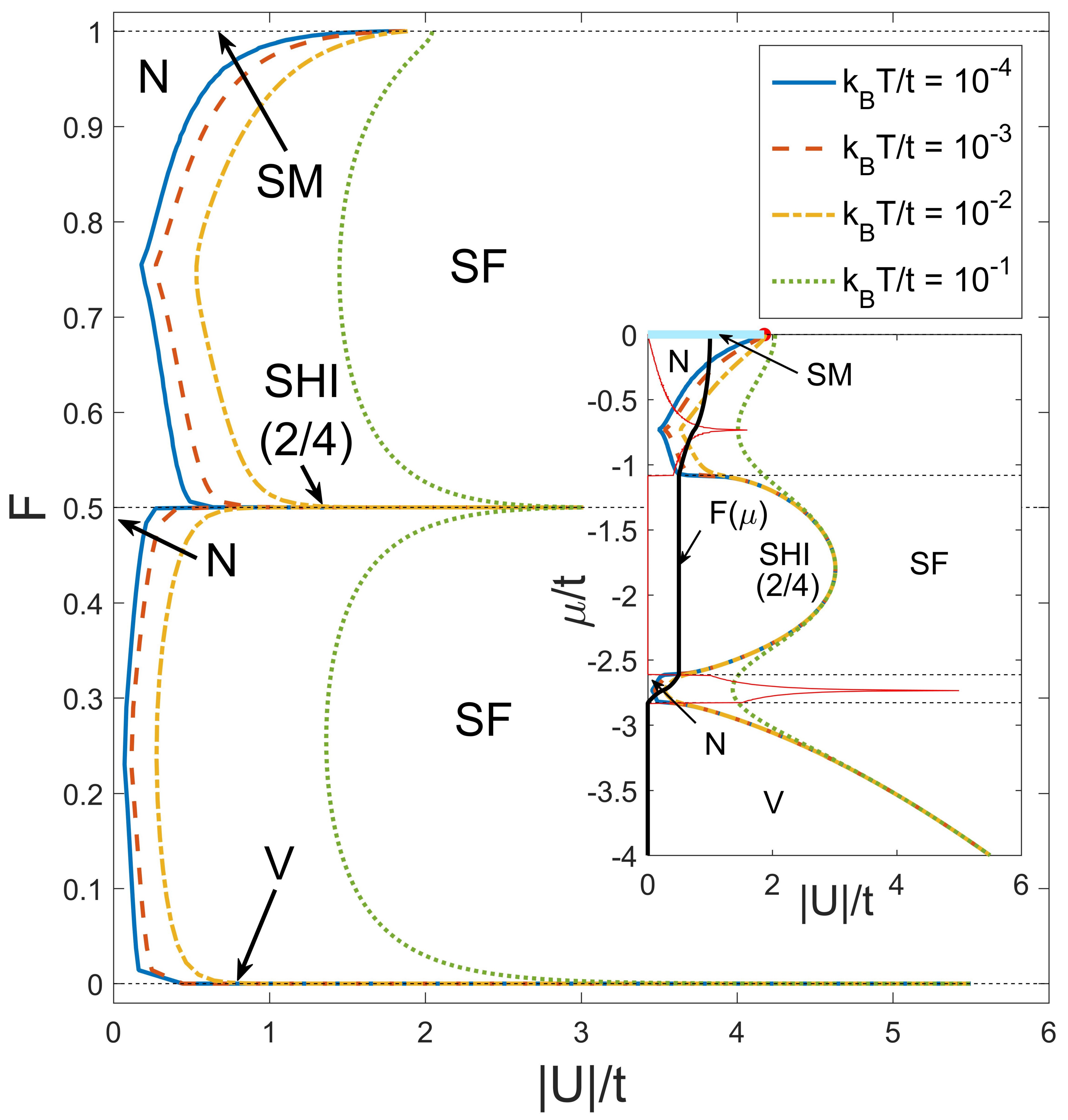}
\includegraphics[scale=0.55]{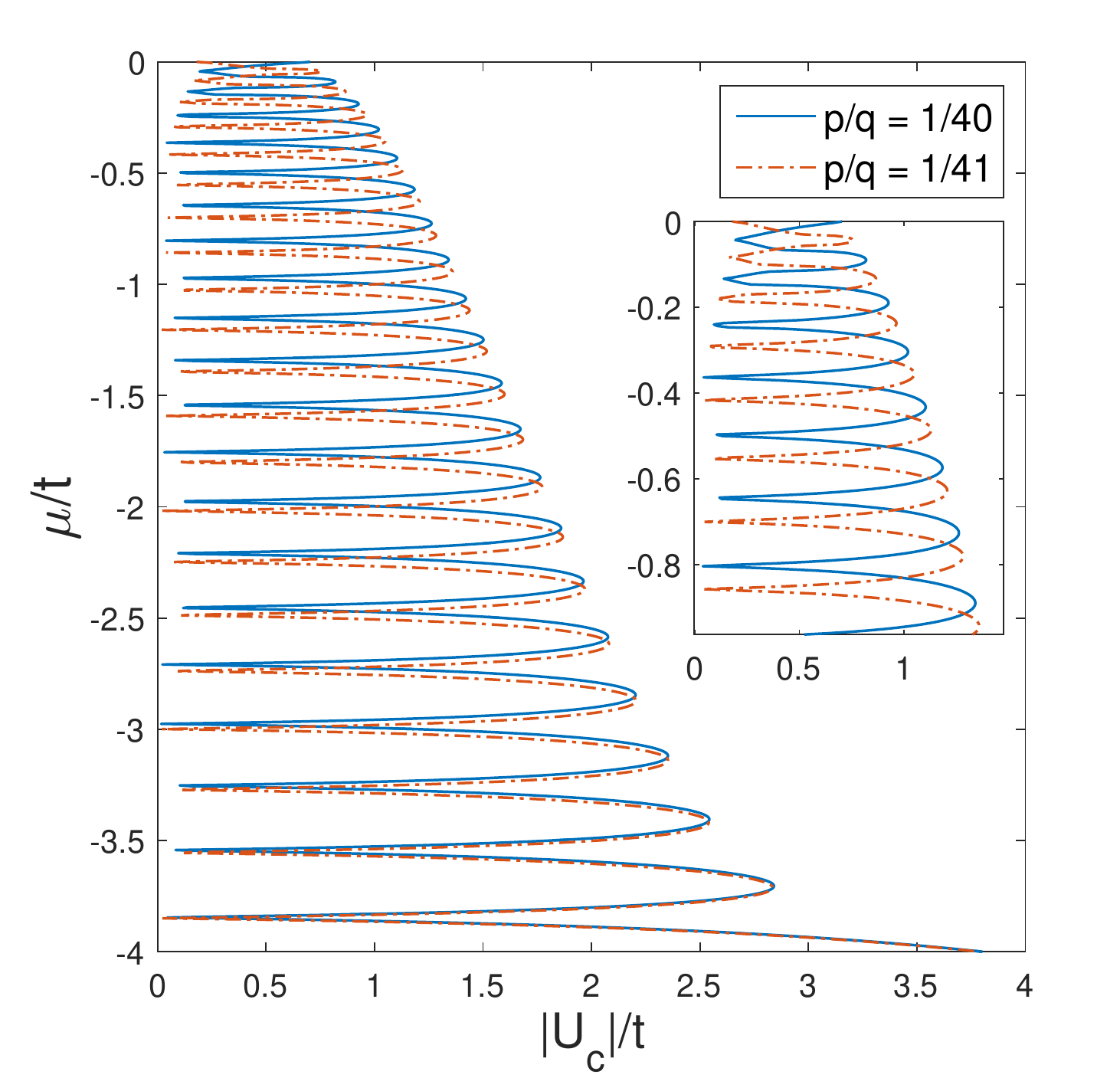}
\caption{Color online)
(Left) Similar to Fig.~1 but $T$ is varied for a fixed $p/q = 1/4$.
(Right) Similar to Figs.~1 and~2 but for large-$q$ denominators.
\label{largeq}
}
\end{figure}
\begin{figure}[htbp]
\includegraphics[scale=0.4]{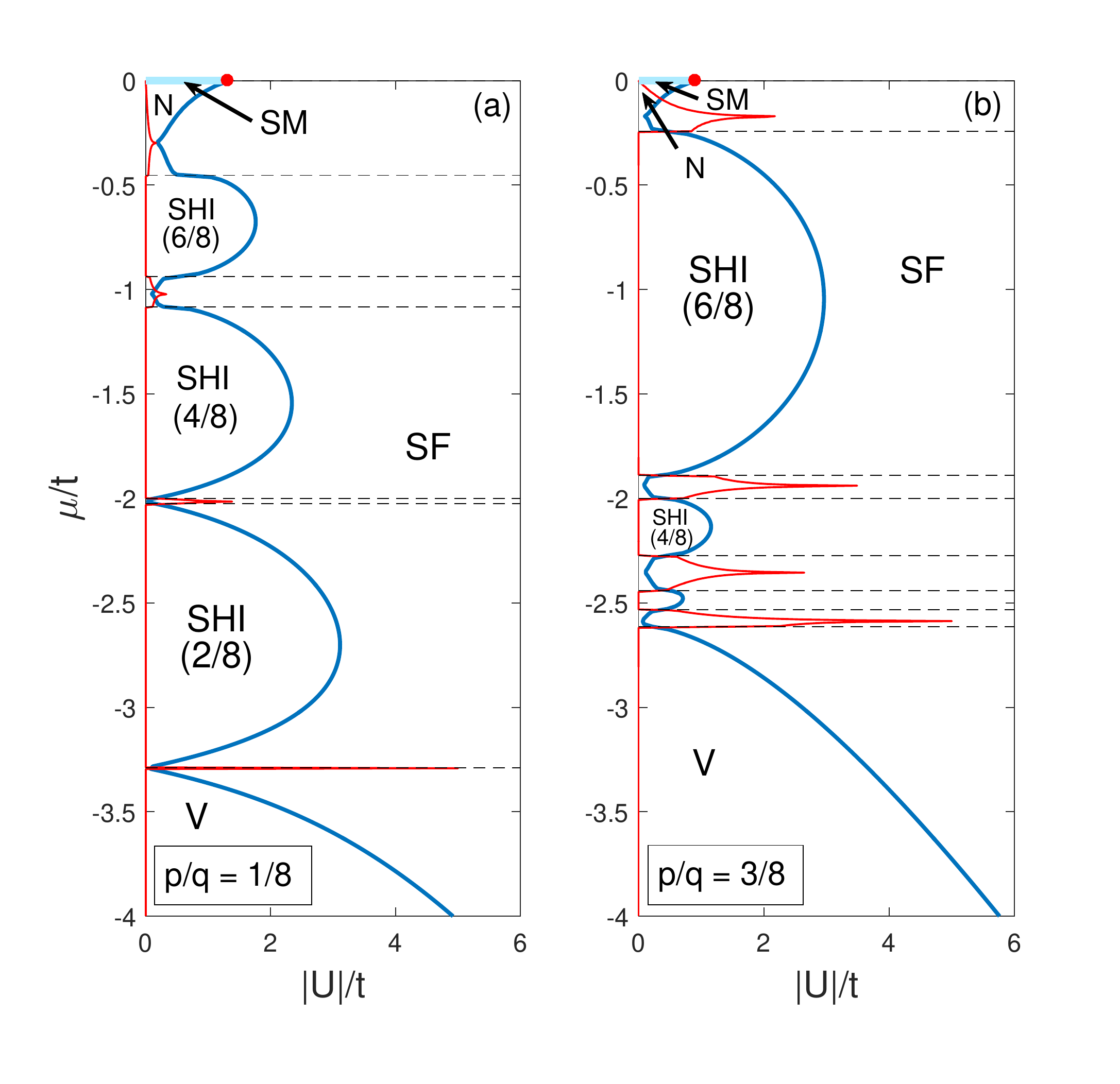}
\includegraphics[scale=0.4]{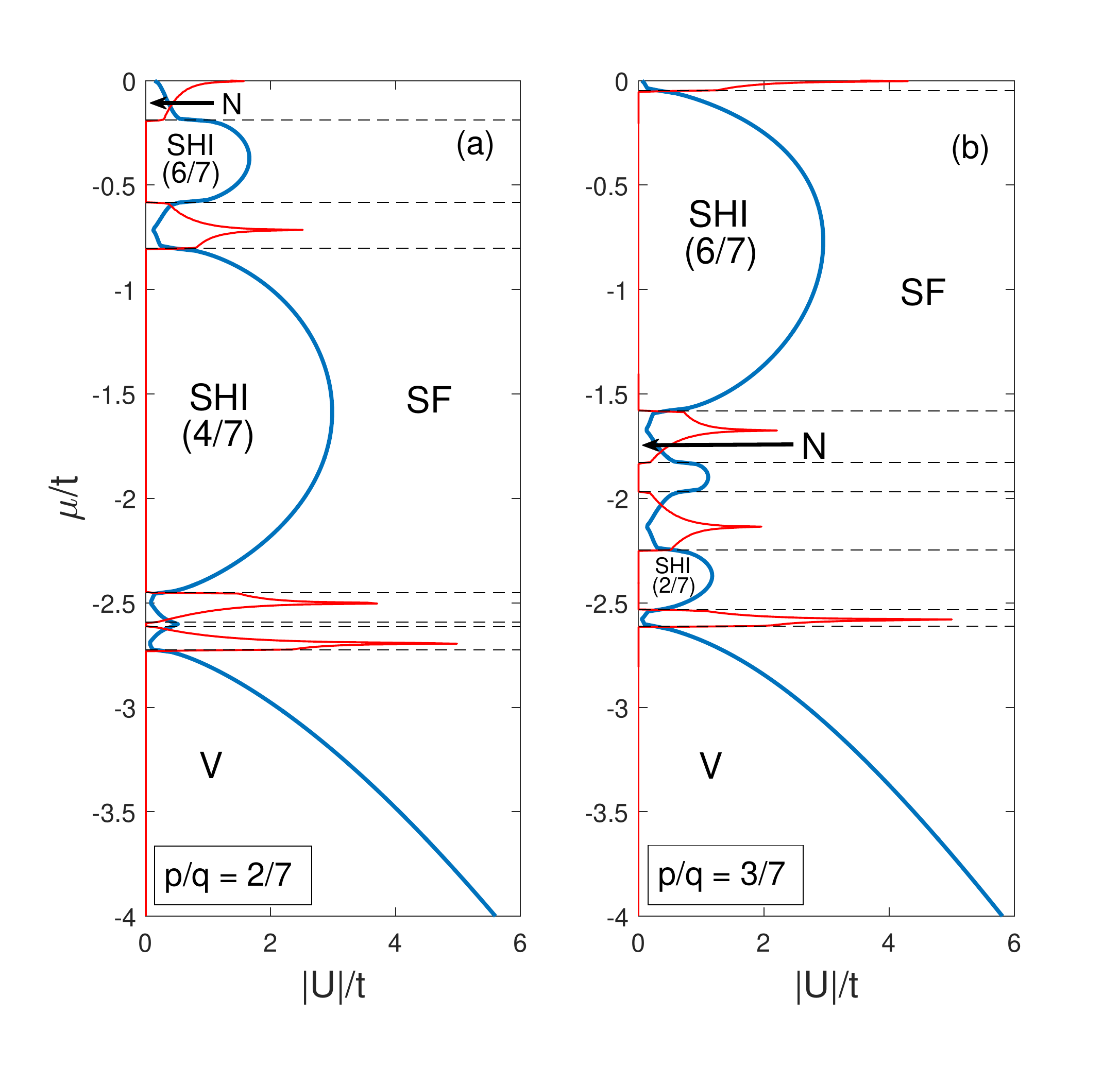}
\caption{Color online)
(Left) Similar to Fig.~1 but for $q = 8$ with (a) $p = 1$ and (b) $p = 3$.  
(Right) Similar to Fig.~2 but for $q = 7$ with (a) $p = 2$ and (b) $p = 3$. 
\label{p1q78}
}
\end{figure}
\end{widetext}


\begin{thebibliography}{99}
\bibitem{BdG} We also confirm this observation with the exact diagonalization of the mean-field
Hamiltonian on a real-space lattice via the BdG prescription~\cite{Iskin}.
\end{thebibliography}
\end{document}